\documentclass[twocolumn,twoside,preprintnumbers,amsmath,amssymb,showkeys]{revtex4}

\usepackage{epsfig}

\usepackage{graphicx}

\usepackage{fancyhdr}
\usepackage{pslatex}

\begin{document}
\title {Shapes and Sizes from Non-Identical-Particle Correlations}

\author{Scott Pratt}

\affiliation{Dept. of Physics and Astronomy,
Michigan State University\\
East Lansing, MI 48824~~USA}

\begin{abstract}
I review the prospects for measuring source characteristics from correlations other than those involving identical pions. Correlations generated from Coulomb and strong interactions are shown to provide remarkable resolving power for determining three-dimensional information, in some cases accessing more detail than can be represented by Gaussian fits.
\keywords{correlation,interferometry,imaging}
\end{abstract}
\pacs{25.75.-q, 25.75.Gz, 25.70.Pq}

\vskip -1.35cm

\maketitle

\thispagestyle{fancy}

\setcounter{page}{1}

\bigskip

\section{Introduction and Theory}

As is well-known to the participants of this conference, identical-particle correlations provide insight into the space-time structure of the asymptotic phase-space density \cite{Lisa:2005dd}. The experimentally measured two-particle correlation function $C({\bf P},{\bf q})$ is related to the source function ${\cal S}({\bf P},{\bf r})$ through a simple Fourier transform,
\begin{equation}
\label{eq:hbt}
{\cal R}_{\bf P}({\bf q})\equiv C_{\bf P}({\bf q})-1
=\int d^3r {\cal S}_{\bf P}({\bf r})\cos(q\cdot r).
\end{equation}
Here, ${\bf P}$ is the total momentum and ${\bf q}=({\bf p}_1-{\bf p_2})/2$ is the relative momentum of the pair. The source function is defined by the outgoing phase space density,
\begin{equation}
{\cal S}_{\bf P}({\bf r})=\frac{\int d^3r_1 d^3r_2 \delta({\bf r}_1-{\bf r}_2-{\bf r})
f({\bf P}/2,{\bf r}_1,t)f({\bf P}/2,{\bf r}_2,t)}
{\int d^3r_1 d^3r_2 f({\bf P}/2,{\bf r}_1,t)f({\bf P}/2,{\bf r}_2,t)},
\end{equation}
where all quantities are calculated in the rest frame of the pair, where ${\bf P}=0$. As long as the time $t$ is beyond any time for which particles are emitted, ${\cal S}$ is independent of $t$ since both particles are moving with identical velocity, which is zero in this frame. The source function physically represents the normalized probability for two particles of the same velocity to be separated by a distance ${\bf r}$ in their asymptotic state. Although the source function only measures a distribution of relative distances, it still provides invaluable insight into the dynamics of a collision. For instance, long lived sources will lead to large separations along the direction of the pair's velocity.

Determining $S$ from the experimentally measured correlation function in Eq. (\ref{eq:hbt}) is simple as one needs only Fourier transform ${\cal R}({\bf q})$. Since the total momentum ${\bf P}$ is not involved in the convolution, I will suppress it in the notation for the remainder of the paper. The ability of correlations of non-identical particles to reproduce source functions is less appreciated by the community, despite the fact that non-identical particle correlations offer similar resolving power to identical-particle correlations in many instances \cite{Pratt:2003ar}. Non-identical-particle correlations are driven by the strong and Coulomb interactions. After making similar, though not identical, assumptions as what were used for Eq. (\ref{eq:hbt}), the corresponding expression for the correlations are:
\begin{equation}
\label{eq:koonin}
{\cal R}({\bf q})=\int d^3r {\cal S}({\bf r})\left(|\phi({\bf q},{\bf r})|^2-1\right).
\end{equation}
Again, the label ${\bf P}$ is suppressed. If one views ${\cal R}({\bf q})$ as a vector, where each bin in ${\bf q}$ represents a component of the vector, and ${\cal S}({\bf r})$ as a vector where each bin in ${\bf r}$ represents a component, determining $S$ from $R$ becomes a matrix inversion problem, where $|\phi|^2-1$ is the matrix with labels ${\bf q}$ and ${\bf r}$. For the rest of the paper I will refer to the matrix as a kernel, ${\cal K}({\bf q},{\bf r})=|\phi|^2-1$.

The goal of this talk is to understand the resolving power of the kernel. When viewed as a matrix, the kernel ${\cal K}({\bf q},{\bf r})$ is highly singular, i.e., it is not possible to uniquely determine every value of ${\cal S}({\bf r})$ from even perfect knowledge of ${\cal R}({\bf q})$. The information made accessible through the kernel depends on the unique properties of the wave function for a particular particle pair. For instance, the $p\Lambda$ wave function is driven only by the strong interaction, which provides a large peak in ${\cal R}({\bf q})$ at small ${\bf q}$. In contrast $pK^+$ correlations are mainly driven by the Coulomb interaction, which tends to wash out the effects of the strong interaction, but provides resolving power through the Coulomb interaction itself. Before discussing the kernels I first introduce the concept of expanding the problem in spherical harmonics or Cartesian harmonics in the next section. In the subsequent sections, I review the general principles determining the resolving power of the kernel for both strong and Coulomb interactions. The resolving power for extracting non-Gaussian features a $pp$ source function is then presented as an example. 

\section{The Resolving Power of Strong and Coulomb Interactions}

Before launching into a detailed discussion of the resolving power of the kernel, ${\cal K}({\bf q},{\bf r})=|\phi|^2-1$, I project the correlation and source function with spherical and Cartesian harmonics \cite{Danielewicz:2005qh}. The convenience of this decomposition derives from the fact that the specific projections of the correlations function, ${\cal R}_{\ell,m}(q)$ for spherical harmonics or ${\cal R}_{\ell_x,\ell_y,\ell_z}$ for Cartesian harmonics, are related to the same projection for the source function,
\begin{eqnarray}
{\cal R}_{\ell m}(q)&=&4\pi \int r^2dr~ {\cal S}_{\ell,m}(r){\cal K}_{\ell}(q,r),\\
\nonumber
{\cal R}_{\vec{\ell}}(q)&=&4\pi \int r^2dr~ {\cal S}_{\vec{\ell}}(r){\cal K}_{\ell}(q,r).
\end{eqnarray}
For Cartesian harmonics, $\ell=\ell_x+\ell_y+\ell_z$. The kernel is the same for both expressions,
\begin{equation}
{\cal K}_{\ell}(q,r)\equiv\frac{1}{2} \int d\cos\theta_{qr}
\left[\left|\phi(q,r,\cos\theta_{qr})\right|^2-1\right] P_\ell(\cos\theta_{qr}).
\end{equation}
The projections reduce angular information $F(\Omega)$ into a set of coefficients, $F_{\ell m}$ or $F_{\vec{\ell}}$.

The projections and inverses for spherical harmonics are defined as:
\begin{eqnarray}
F_{\ell m}&\equiv&(4\pi)^{-1/2}\int d\Omega F(\Omega)Y_{\ell m}(\Omega),\\
\nonumber
F(\Omega)&\equiv&(4\pi)^{1/2}\sum_{\ell,m}F_{\ell m}Y_{\ell m}(\Omega).
\end{eqnarray}
The $(2\ell+1)$ complex coefficients obey the symmetry $F_{\ell m}=(-1)^m F^*_{\ell -m}$, so that the angular information for each $\ell$ is represented by $(2\ell+1)$ independent real numbers.

For Cartesian harmonics, the projections are defined as:
\begin{eqnarray}
\label{eq:harmexpansion}
F_{\vec{\ell}}&=&\frac{(2\ell+1)!!}{\ell!}\int\frac{d\Omega}{4\pi}
F(\Omega)A_{\vec{\ell}}(\Omega),\\
\nonumber
F(\Omega)&=&\sum_{\vec{\ell}}\frac{\ell!}{\ell_x!\ell_y!\ell_z!}F_{\vec{\ell}}
\hat{n}_x^{\ell_x} \hat{n}_y^{\ell_y} \hat{n}_z^{\ell_z}
\end{eqnarray}
The Cartesian harmonic function $A_{\vec{\ell}}(\Omega)$ represent products of unit vectors, $\hat{n}_x^{\ell_x}\hat{n}_y^{\ell_y}\hat{n}_z^{\ell_z}$, which then have lower values of $\ell$ projected away. Table \ref{table:cartesian} shows the functions for $\ell\le 4$. Unlike the spherical harmonics, all these functions are real. For a given $\ell$, there are $(\ell+1)(\ell+2)/2$ different combinations of $(\ell_x,\ell_y,\ell_z)$ that sum to $\ell$. However, not all coefficients are independent. Since any part of the function from which one can factor $n_x^2+n_y^2+n_z^2=1$ gives a function of lower projection $\ell$,  the functions $A_{\vec{\ell}}$ and the coefficients $F_{\vec{\ell}}$ satisfy the ``tracelessness'' constraint,
\begin{equation}
F_{\ell_x+2,\ell_y,\ell_z}+F_{\ell_x,\ell_y+2,\ell_z}+F_{\ell_x,\ell_y,\ell_z+2}=0.
\end{equation}
With this constraint, all the coefficients for a given $\ell$ can be generated unambiguously from knowing the $(2\ell+1)$ coefficients for $\ell_x=0$ or 1. Thus, both expansions, with spherical harmonics or Cartesian harmonics, can be represented by $(2\ell+1)$ real numbers for each $\ell$.
\begin{table}
\caption{\label{table:cartesian} Cartesian harmonics for $\ell<=4$. Other
harmonics can be found by swapping indices on both sides of the
equation, e.g., $x\leftrightarrow y$. For example, given ${\mathcal
A}_{210}=n_x^2n_y-n_y/5$, swapping $y\leftrightarrow z$ gives ${\mathcal
A}_{201}=n_x^2n_z-n_z/5$.}
\begin{tabular}{|c|c|}\hline
${\mathcal A}^{(1)}_{100}=n_x$ 
& ${\mathcal A}^{(3)}_{111}=n_xn_yn_z$ \\
${\mathcal A}^{(2)}_{200}=n_x^2-(1/3)$ 
& $A^{(4)}_{400}=n_x^4-(6/7)n_x^2+(3/35)$ \\
${\mathcal A}^{(2)}_{110}=n_xn_y$ 
& ${\mathcal A}^{(4)}_{310}=n_x^3n_y-(3/7)n_xn_y$\\
$A^{(3)}_{300}=n_x^3-(3/5)n_x$ 
&  ${\mathcal A}^{(4)}_{220}=n_x^2n_y^2-(1/7)n_x^2-(1/7)n_y^2+(1/35)$\\
${\mathcal A}^{(3)}_{210}=n_x^2n_y-(1/5)n_y$ 
& ${\mathcal A}^{(4)}_{211}=n_x^2n_yn_z-(1/7)n_yn_z$\\ \hline
\end{tabular}
\end{table}

The two choices of basis functions for the projections are equivalent, and it is straight forward to convert between one set of coefficients and the other. The main advantage of Cartesian harmonics is that it is easier to identify the coefficient with the distortion of the shape as can be seen by the expansions in Eq. (\ref{eq:harmexpansion}).

\section{The Resolving Power of Identical-Particle Statistics}

The kernels for identical particles are particularly simple,
\begin{eqnarray}
{\cal K}({\bf q},{\bf r})&=&\cos({\bf q}\cdot{\bf r}),\\
\nonumber
{\cal K}_{\ell}(q,r)&=&j_\ell(qr).
\end{eqnarray}
Inverting ${\cal R}(q)$ to find ${\cal S}(r)$ involves a simple Fourier transform. Roughly speaking, the correlation structure of level $q$ reveals information about components of the source function of scale $R\sim 1/q$. Shape characteristic are also easy to invert using the kernel for specific $\ell$. Since $j_\ell(qr)$ behaves as $(qr)^\ell$, information about the distortions of a specific $\ell$ require looking at a range of $q$ for which $qR\gtrsim \ell$. The fact that the kernel does not provide resolving power for small $r$ is not usually a problem, since the source function ${\cal S}_\ell(r)$ also tends to   vanish for $(r/R)<<\ell$.

\section{The Resolving Power of Strong Interactions}

First, I consider the information available by analyzing the kernel for $\ell=0$. For $q>>1/R$, where $R$ is a characteristic scale of the source, $|\phi(q,r)|^2$ differs significantly from unity only at small $r$, and the kernel can be approximated by a delta function. The correlation function is then only sensitive to the probability that the two particles have $r=0$, with a resolving power determined by the overall density of states which is given by the scattering phase shifts,
\begin{eqnarray}
\label{eq:strongkernel_ell0}
{\cal R}_{\ell=0}(q)&=&\frac{\Delta dn/dq}{4\pi q^2/[(2\pi\hbar)^3]}{\cal S}({\bf r}=0),\\
\Delta dn/dq&=&\frac{1}{\pi}\sum_\ell (2\ell+1)\frac{d\delta_\ell}{dq}.
\end{eqnarray}
The sensitivity in $q$ is determined by the form of $d\delta/dq$, rather than by details of the shape.

For $q\lesssim 1/R$, where $R$ is a characteristic scale of the source, ${\cal R}_{\ell=0}(q)$ becomes sensitive to more details of ${\cal S}_{\ell=0}(r)$, not just the value at $r=0$. At that point, the resolving power depends on the unique shape of the  wave function, which is also given by the phase shifts. For $q<<1/R$, the kernel is determined solely by the $\ell=0$ phase shifts. As long as the range of the interaction is short compared to the source size (true for heavy ion collisions), one can then approximate the wave function by a distorted plane wave. If only the $\ell=0$ pieces are kept, 
\begin{eqnarray}
{\cal K}({\bf q},{\bf r})\approx
\left|e^{iq\cdot r}+\frac{\sin (qr+\delta_{\ell=0})}{qr}-\frac{\sin qr}{qr}\right|^2-1\\
\nonumber
{\cal K}({\bf q}\rightarrow 0,{\bf r})\approx \frac{-2ar+a^2}{r^2},
\end{eqnarray}
where $a$ is the scattering length, $\delta_{\ell=0}=-qa$. Thus, one can crudely state that the ${\cal R}_{\ell=0}(q\rightarrow 0)$ provides information about the $\langle 1/r^2\rangle$ and $\langle 1/r\rangle$ moments of the source function, while ${\cal R}_{\ell=0}(q>>1/R)$ provides information about ${\cal S}_{\ell=0}(r\rightarrow 0)$. Looking at $d{\cal R}_{\ell=0}/dq^2$ at $q=0$ will provide insight into other higher moments such as $\langle 1\rangle$ and $\langle r\rangle$. For higher moments the constant of proportionality depends on the $\ell=1$ phase shifts and on the next higher expansion of the $\ell=0$ phase shift, which is determined by the effective range. 

Gaussian parameterizations of ${\cal S}_{\ell=0}(r)$ involve two parameters, a size $R_G$, and a ``coherence'' parameter $\lambda$ which represents the normalization of ${\cal S}$ which can differ from unity since some of the particles might originate from effectively infinite distances due to long-lived decays. Any two pieces of information about ${\cal S}_{\ell=0}(r)$, such as the value at $r=0$ and a moment, are sufficient to determine the two Gaussian parameters. As explained in the previous paragraph, the kernel can, in principle, provide more information. However, the ability to extract such non-Gaussian details are constrained by the strength and form of the scattering phase shifts and by statistical and systematic errors in the data.

The kernel for $\ell>0$ allows one to determine properties of the source shape. Even though the scattering phase shifts are usually important only for a few small values of $\ell$, ${\cal K}_{\ell}(q,r)$ has strength for all $\ell$. This  sensitivity can be explained by pointing out that for large $qr$, which is the only region one can analyze higher $\ell$, one can view the problem classically. The kernel is then largely determined by the shadowing. In the large $qr$ limit the kernel can be determined solely from the geometry of classical trajectories,
\begin{equation}
\label{eq:strongkernel_bigell}
\left.{\cal K}({\bf q},{\bf r})\right|_{r\rightarrow\infty}
=-\frac{1}{r^2}\left\{
-\sigma\delta(-\Omega_{qr})+\frac{d\sigma}{d\Omega}
\right\}.
\end{equation}
Here, the first term represents the shadowing due to particles scattering away from the direction $\hat{q}\cdot\hat{r}=-1$, while the second term represents the addition due to scattered particles. Since the delta function has contributions to all $\ell$, the kernel from strong interactions provides resolving power at all $\ell$ for larger $r$. In practice, the power falls off at large $\ell$ since quantum considerations destroy ${\cal K}_\ell$ for $qr\lesssim \ell$.

The $\ell=0$ kernel shown in Eq. (\ref{eq:strongkernel_ell0}) and the $\ell>0$ kernel shown in Eq. (\ref{eq:strongkernel_bigell}) tend to be of similar strength for most circumstances. Thus, for most interactions, it can be stated that if one has sufficient statistics to measure the size, one also has the ability to determine shape.

\section{The Resolving Power of Coulomb Interactions}

Coulomb interactions also provide leverage for determining source sizes. The limitations of the Coulomb interaction derives from the inherently weak couplings which make correlations relatively small except at small $q$. At small $q$, $|\phi|^2\rightarrow 0$, unless is so large as to be outside the range of the Coulomb interaction, $e^2/r << q^2/2\mu$. Thus, all pairs are emitted within a radius, $r<\mu e^2/q_{\rm r}^2$, where $q_r$ is the momentum resolution, contribute zero to the correlation function, or equivalently $-1$ to the kernel. For most experiments the resolution is of the order of 1 MeV/$c$ and only those pairs where at least one member originates from a weak decay are uncorrelated. Thus, the intercept of the correlation function at $q=0$ gives
\begin{equation}
C(q\rightarrow 0)=(1-f_{\rm l.l.})^2,
\end{equation}
where $f_{\rm l.l.}$ is the fraction of particles from long-lived decays.

For large $q$, one can use the classical limit of the kernel given by calculating classical Coulomb trajectories,
\begin{eqnarray}
\label{eq:coulapprox}
|\phi(q,r,\cos\theta_{qr})|^2&\rightarrow& \frac{d^3q_0}{d^3q}\\
\nonumber
&&\hspace*{-50pt}=\frac{1+\cos\theta_{qr}-x}
{\sqrt{(1+\cos\theta_{qr}-x)^2-x^2}}
\Theta(1+\cos\theta_{qr}-2x).
\end{eqnarray}
Here, $x$ is the ratio of the Coulomb and final-state kinetic energy,
\begin{equation}
x\equiv\frac{2\mu e^2}{q^2r}.
\end{equation}
Since $e^2=1/137$ is small, $x$ is small unless $q$ is quite small. This allows the kernel to be approximated as
\begin{eqnarray}
{\cal K}({\bf q},{\bf r})|_{x\rightarrow 0}&\approx&
-\frac{x}{2}\delta(1+\cos(\theta_{qr}),\\
\nonumber
{\cal K}_{\ell}(q,r)|_{x\rightarrow 0}&\approx&(-1)^\ell \frac{x}{2}.
\end{eqnarray}
In addition to $x$ being small, this approximation also relies on the classical approximation, $qr>>\ell$.

The Coulomb interaction thus provides excellent resolving power for all $\ell$, provided that $x$ is not too small for the relevant region $qR\gtrsim 1$, which can equivalently be stated that $\mu e^2 R$ should not be too small, or the source size should not be too much smaller than the Bohr radius, which is 390 fm for $\pi\pi$ but only 58 fm for $pp$. Thus, one expects reasonable resolving power for $pK$ or heavier pairs, whereas pairs involving pions offer weak resolving power through the Coulomb force. The Coulomb interaction becomes especially useful if light nuclei can be used, as in addition to the increase in the reduced mass $\mu$, the increased factor of charges, $e^2\rightarrow Z_1Z_2e^2$, bolsters the resolving power.

For large $r$ the Coulomb kernel falls as $(q^2r)^{-1}$ for large $q$. Thus, the tail of the correlation function provides a measure of the moment $\langle 1/r\rangle$ for the source function. By analyzing the correlation function for its variance from the $1/q^2$ behavior for decreasing $q$, one can, in principle, ascertain other facets of the source function.

\section{Resolving Power in the Real World}

Imaging has become a popular term for describing the inversion of ${\cal R}(q)$ to determine ${\cal S}(q)$. Particular attention has been given to the $\ell=0$ components by Danielewicz and Brown by fitting to splines \cite{Brown:2000aj,Brown:1999ka,Brown:1997ku}, where non-Gaussian aspects of source functions have been observed. The spline-fitting routines are themselves a fit to a constrained functional form, and should not be mistaken for a fully flexible accounting for all possible forms for the source function. Our goal here is to gain a better feel for the sensitivity to non-Gaussian features of the source one might attain in real-world analyses.

First, I will consider the simple case of $\ell=0$ sources. I pick the form
\begin{equation}
\label{eq:GX}
{\cal S}_{\ell=0}(r)
=\lambda (1-X_{\rm frac})\frac{e^{-r^2/2R^2}}{Z_G}
+\lambda X_{\rm frac}\frac{e^{-\sqrt{r^2/X^2+4R^4/X^4}}}{Z_X}.
\end{equation}
Here, the normalization constant ensures that each source would integrate to unity when $\lambda=1$. The four parameters in this fit are:
\begin{itemize}
\item $\lambda$, the ``coherence factor''.\vspace*{-8pt}
\item $R_G$, a Gaussian source size.\vspace*{-8pt}
\item $X$, an exponential size describing the source function at large $r$.\vspace*{-8pt}
\item $X_{\rm frac}$, the fraction of the source described by the exponential.
\end{itemize}
The form of the argument of the second term was chosen to force the curvature at the origin to match that of the Gaussian.

\begin{figure}[hbt]
\centerline{\includegraphics[width=0.4\textwidth]{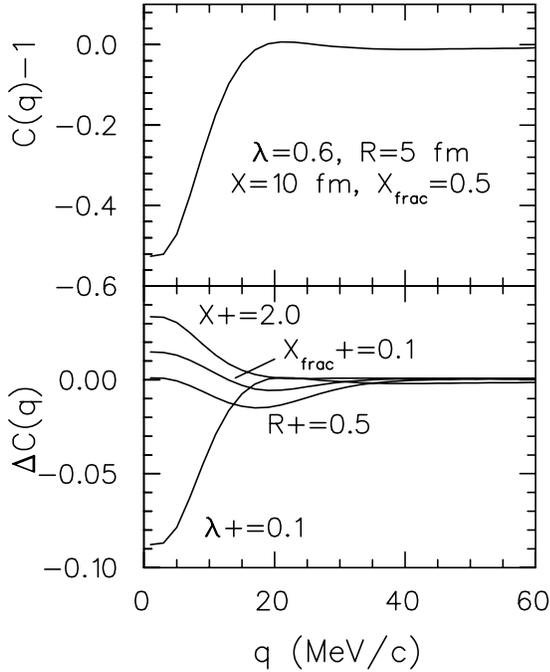}}
\caption{\label{fig:GX}
A one-dimensional $pp$ correlation function for the parameters shown in Eq. (\ref{eq:GX}) is displayed in the upper panel, while changes corresponding to the labeled changes of the four parameters are shown in the lower panel. For most experiments, only three of the four parameters would be uniquely determined.
}
\end{figure}

The upper panel of Fig. \ref{fig:GX} shows a $pp$ correlation function calculated for the form above with the parameters: $\lambda=0.6$, $R_G=5$ fm, $X=10$ fm, $X_{\rm frac}=0.5$. In the lower panel, the change of the correlation function is illustrated for various changes in the four parameters above. Visually, it is apparent that any one of the four changes can be pretty well approximated by making linear combinations of the other three. For instance, the effects of increasing the exponential length $X$ by 2 fm can be approximately reproduced by a combination of reducing $\lambda$ combined with small changes to the remaining two parameters.

The ability of an experimental analysis to determine all four parameters in Eq. (\ref{eq:GX}) depends on details of the particular interaction and the experimental accuracy and resolution. To more rigorously determine the degree to which parameters could be experimentally constrained, theoretical correlation functions were binned, and random errors were then added. For errors of the order of what might be expected at RHIC, it did not seem possible to uniquely determine all four parameters. However, extracting three parameters was usually quite robust. Since a Gaussian parameterization uses only two parameters, our exercise demonstrates that non-Gaussian features can be extracted, but that determining both the weight $X_{\rm frac}$ and the exponential scale $X$ will probably be outside the resolution of the experiments. However, if one feels confident in the scale $X$, e.g., one might believe it comes principally from $\omega$ decays, our investigations suggest that it is quite possible to confidently determine the relative weight $X_{\rm frac}$. Most of the pairs that were investigated appeared to accommodate three-parameter analyses.

\begin{figure}[hbt]
\centerline{\includegraphics[width=0.38\textwidth]{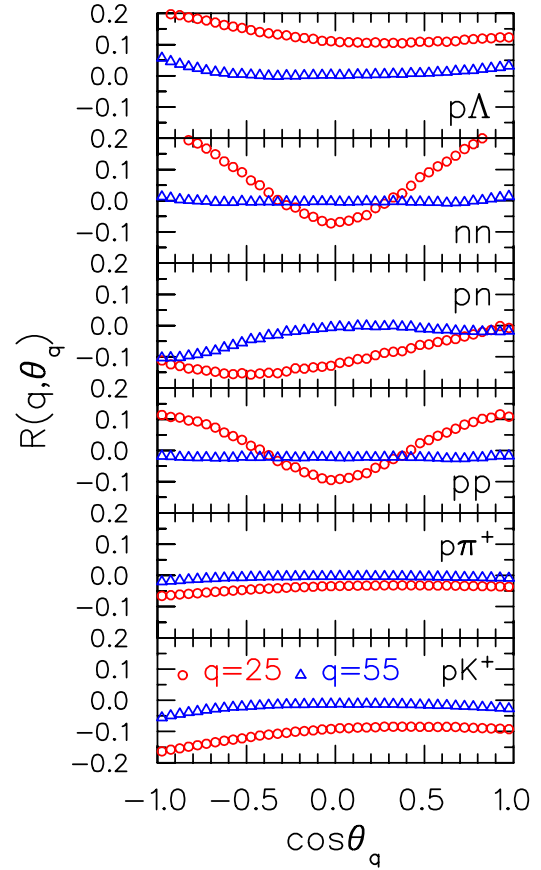}}
\caption{\label{fig:c_vs_ctheta}
Correlations for two relative momenta are shown as a function of the angle with respect to the $z$ axis for a Gaussian source of radius $R_x=R_y=4$, $R_z=8$ fm. The angular variations for $q=55$ MeV/$c$ illustrate significant resolving power  for experimentally determining shapes. 
}
\end{figure}
\begin{figure}[hbt]
\centerline{\includegraphics[width=0.47\textwidth]{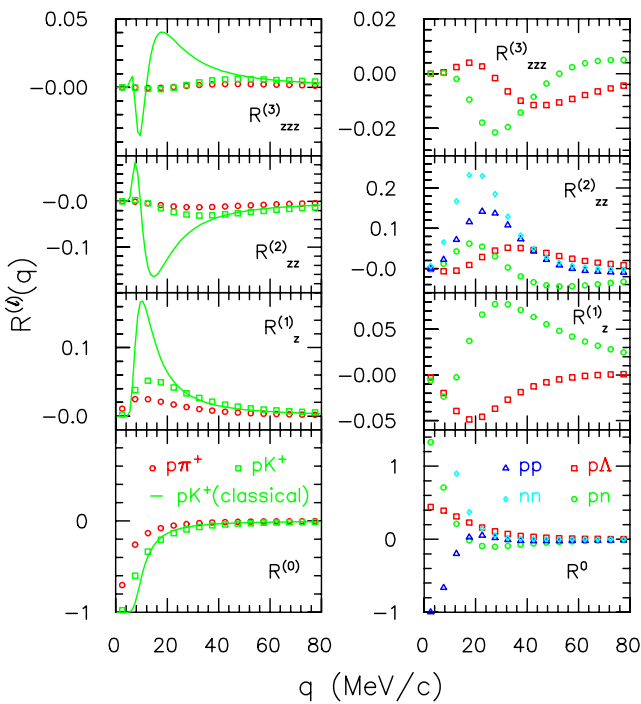}}
\caption{\label{fig:c_gauss}
Angular projections of correlations as a function of $q$ for a Gaussian source of dimensions $R_x=R_y=4$ fm, $R_z=8$ fm. For non-identical particles, the center of the Gaussian was also moved by 4 fm in the $z$ direction. Projections are for Cartesian harmonics, with $\ell=0$ (lower panels), $\ell_x=\ell_y=0,\ell_z=1,2,3$, subsequently higher panels.
}
\end{figure}

As shown in the previous section, insight into the three-dimensional shape of sources can come from all types of interactions. This is illustrated by the plotting of several angular correlations for a non-spherical source in Fig. \ref{fig:c_vs_ctheta}. The Gaussian source has dimensions of 4 fm in the $x$ and $y$ directions and 8 fm in the $z$ direction. For non-identical particles, the center of the Gaussian was moved by 4 fm to simulate one species being emitted ahead or behind of the other \cite{Lednicky:1995vk,gelderloos,Voloshin:1997jh}. The correlation function is plotted at two values of $q$ as a function of the angle with respect to the $z$ axis. The fact that the values are not constant, and that their variation with angle is similar to their average strength, illustrates the strong angular resolving power provided by all the pairs. In all cases, their is little sensitivity to shape at small $q$. This can be understood from considering the uncertainty principle that says that if $qR\lesssim 1$, there is an inherent vagueness as to the direction of the relative momentum. However, at a relative momentum of 55 MeV/$c$, the angular sensitivity is significant for all cases. The pairs involving two baryons tended to be especially strong.
To illustrate the same information with angular projections, Fig. \ref{fig:c_gauss} shows projections in terms of Cartesian harmonics for all $\ell\le 3$. 

\begin{figure}[hbt]
\centerline{\includegraphics[width=0.38\textwidth]{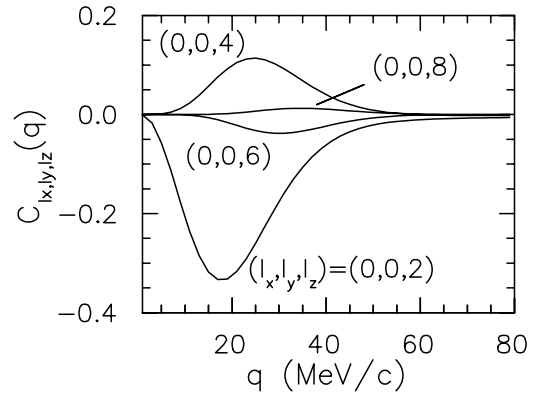}}
\caption{\label{fig:cartharm_gauss_cf_pipi}
Angular projections of identical-pion correlations as a function of $q$ for a Gaussian source of dimensions $R_x=R_y=4$ fm, $R_z=8$ fm. Since the correlations are stronger for this case, larger values of $\ell$ can be explored.
}
\end{figure}

As a point of comparison, I present identical-pion correlations for the same source, ($R_x=R_y=4$ fm, $R_z=8$ fm), but without the offset of the center due to the fact that there can be no offset with identical particles. As can be seen in Fig. \ref{fig:cartharm_gauss_cf_pipi}, the correlations are much larger than those seen for any of the examples from Fig. \ref{fig:c_gauss}. The stronger resolving power derives from the simple fact that all the correlations, including those for $\ell=0$, are several times larger. For identical particles, there are no odd-numbered harmonics. However, as shown in Fig. \ref{fig:cartharm_gauss_cf_pipi}, it should be feasible to measure harmonics to rather large $\ell$.

In addition to the calculations for Gaussians shown here, I have evaluated angular projections for correlations from blast wave models. From our experience thus far I can make the following rough statements regarding the prospects for discerning shapes.
\begin{itemize}
\item For identical pion or identical kaon correlations, many values of $\ell$ are accessible.
\item For $pK$ or $p\pi$ correlations:
\begin{itemize}
\item $\ell=1$ terms are easy
\item $\ell=2$ require 1\% or better accuracy
\item $\ell=3$ are probably prohibitively small
\end{itemize}
\item For baryon-baryon correlations
\begin{itemize}
\item $\ell=1,2$ are easy
\item $\ell=3$ requires 1\% or better accuracy.
\end{itemize}
\end{itemize}

\section{Summary and Prospects}

Identical-pion correlations arguably represent the most discriminating observable at RHIC. A large fraction of the space of possible equations of state have been excluded by theory/experiment comparison, including strong first-order equations of state, and on the opposite extreme, hard pion-gas-like equations of state. Given the paramount importance of these conclusions, it would be enormously important to verify these statements with independent analyses of space-time parameters using other classes of correlations. As a first goal, experimental analyses should work on alternative measurements of $R_{\rm out}/R_{\rm side}/R_{\rm long}$. Additionally, the odd harmonics access new features of the emission geometry. These alternative measurements should be viewed as being more complementary than redundant, as the theoretical basis for Coulomb-induced and strong-interaction based correlations is somewhat different from that used for identical-particle interference. 

Exploiting other classes of shape analyses is inherently difficult for two reasons. First, the strength of the correlations tends to be small. Whereas identical pion correlations are measured in the tens of percent, other classes of correlation tend to be a few percent once they are in the region $q\gtrsim 25$ MeV/$c$, where one can access shape information. Recent data sets from RHIC have amassed sufficient statistics to analyze correlation functions at the sub-one-percent level. The PHENIX experiment, with its excellent particle-identification over a wide range of momenta, is especially well poised to exploit such correlations. STAR has even better statistics for some cases, but the delay in finishing the STAR time-of-flight wall has made it difficult to access some correlations at high statistics. For instance, since correlations are analyzed at small relative velocity (not small relative momentum) $p\pi$ correlations require comparing a $p_t=100$ MeV/$c$ pion to a $700$ MeV/$c$ proton to match velocities. Both particles are at the edge of STAR's acceptance for particle identification. However, analyses will ultimately be constrained by systematic experimental errors, and from competing physical processes, such as flow or jets. All such effects must be handled more carefully when the desired correlation is of order one percent.

The second challenge for analyzing these classes of correlation derives from the  difficulty in disentangling the more complicated and opaque kernels. However, as was shown here, this is by no means insurmountable. Recently developed fitting algorithms, imaging techniques, and angular decompositions, have now been compiled into usable libraries with the CorAL (Correlations Analysis Library) project. An alpha version of the library is already available \cite{coralpha}, and interested parties are encouraged to contact any of the authors, for assistance in using the codes.

\bigskip
\noindent\textbf{Acknowledgments}

\noindent
The generous support of the U.S. Dept. of Energy through grant no. DE-FG02-03ER41259 is gratefully acknowledged.

\end{document}